\documentclass[prd,amsmath,amssymb,nofootinbib,superscriptaddress]{revtex4}
\usepackage{array,mathtools,amssymb,booktabs,multirow,diagbox,hhline,natbib}
\pdfoutput=1

\usepackage{graphicx,graphics,color}% Include figure files
\usepackage{dcolumn}% Align table columns on decimal pointM_\pi
\usepackage{bm}% bold math
\usepackage{bbold}
\usepackage{amsmath}
\usepackage{xcolor}

\allowdisplaybreaks

\newcommand{\mean}[1]{\left\langle{#1}\right\rangle}
\newcommand{\condl}{\mean{\bar q q}_l}
\newcommand{\conds}{\langle \bar s s \rangle}
\begin{document}

\title{Thermal hadron resonances in chiral and $U(1)_A$ restoration}
\author{A. G\'omez Nicola}
\email{gomez@ucm.es}
\affiliation{Universidad Complutense de Madrid, Facultad de Ciencias F\'isicas, Departamento de F\'isica Te\'orica and
IPARCOS. Plaza de las Ciencias 1, 28040 Madrid, Spain}
\author{J. Ruiz de Elvira}
\email{jacobore@ucm.es}
\affiliation{Universidad Complutense de Madrid, Facultad de Ciencias F\'isicas, Departamento de F\'isica Te\'orica and
IPARCOS. Plaza de las Ciencias 1, 28040 Madrid, Spain}
\author{A. Vioque-Rodr\'iguez}
\email{avioque@ucm.es}
\affiliation{Universidad Complutense de Madrid, Facultad de Ciencias F\'isicas, Departamento de F\'isica Te\'orica and
IPARCOS. Plaza de las Ciencias 1, 28040 Madrid, Spain}

\begin{abstract}
We review recent work on thermal resonances and their connection with chiral symmetry and $U(1)_A$ restoration within the QCD phase diagram. In particular, the $f_0(500)$ and $K_0^* (700)$ states generated from $\pi\pi$ and $\pi K$ scattering within Unitarized Chiral Perturbation Theory (ChPT) at finite temperature allow one to describe scalar susceptibilities, which combined with Ward Identities yield interesting conclusions regarding the interplay between chiral and $U(1)_A$ restoration, key to understand the nature of the transition.
\end{abstract}

\maketitle

%\markboth{Angel Gómez Nicola, Jacobo Ruiz de Elvira, Andrea Vioque-Rodríguez}
%%
%{THERMAL HADRON RESONANCES  IN CHIRAL AND $U(1)_A$ RESTORATION}
%%
%\begin{document}
%\title{THERMAL HADRON RESONANCES  IN CHIRAL AND $U(1)_A$ RESTORATION
%\vspace{-6pt}}
%\author{Angel Gómez Nicola, Jacobo Ruiz de Elvira, Andrea Vioque-Rodríguez}
%\address{Universidad Complutense de Madrid, Facultad de Ciencias Físicas, Departamento de Física Teórica and
%IPARCOS, Plaza de las Ciencias 1, 28040 Madrid, Spain}
%\address{ }
%\author{ }
%\address{ }
%\author{ }
%\address{ }
%\author{ }
%\address{ }
%\maketitle
%\recibido{day month year}{day month year
%\vspace{-12pt}}
%\begin{abstract}
%%\vspace{1em} 
%We review recent work on thermal resonances and their connection with chiral symmetry and $U(1)_A$ restoration within the QCD phase diagram. In particular, the $f_0(500)$ and $K_0^* (700)$ states generated from $\pi\pi$ and $\pi K$ scattering within Unitarized Chiral Perturbation Theory (ChPT) at finite temperature allow one to describe scalar susceptibilities, which combined with Ward Identities yield interesting conclusions regarding the interplay between chiral and $U(1)_A$ restoration, key to understand the nature of the transition.
%% \vspace{1em}
%\end{abstract}
%\keys{ Hadron Physics, QCD Phase Diagram, Thermal resonances  \vspace{-4pt}}
%\pacs{   \bf{\textit{11.30.Rd, 11.10.Wx, 12.39.Fe }}    \vspace{-4pt}}
%\begin{multicols}{2}

\section{Introduction}

The analysis and understanding of the QCD Phase Diagram is one of the major research lines in hadronic physics where considerable progress has been reached within recent years, mostly from lattice field theory and theoretical developments based on effective theories \cite{Aoki:2009sc,Borsanyi:2010bp,Bazavov:2011nk,Ratti:2018ksb,Bazavov:2019lgz,GomezNicola:2020yhm,Nicola:2020smo}. The coincidence of  deconfinement and chiral symmetry restoration in the $(T,\mu_B)$ plane of temperature and baryon chemical potential allows one to use genuine chiral-restoring observables as signals of the transition, as we will see here. In fact, although observables related to deconfinement, such as the trace anomaly, or others indicating chiral symmetry breaking, such as the light quark condensate $\condl$, receive contributions from many massive hadronic states and require the implementation of Hadron Resonance Gas (HRG) based approaches \cite{Karsch:2003vd,Huovinen:2009yb,Jankowski:2012ms,Andronic:2012ut,Huovinen:2017ogf}, we will focus here on those  for which an effective theory approach based only on light hadronic degrees of freedom provides  the dominant physical description.  That is the case of observables constructed out of the following quark bilinears in the light channels:

\begin{eqnarray}
I=0&:&  \sigma_l=\bar\psi_l \psi_l \ (S), \ \eta_l=i\bar\psi_l\gamma_5 \psi_l \ (P) \nonumber\\
I=1&:&  \pi^a=i\bar\psi_l\gamma_5\tau^a\psi_l \ (P), \ \delta^a=\bar\psi_l \tau^a \psi_l  \ (S)   \nonumber\\
I=1/2&:& K^b=i\bar\psi  \gamma_5 \lambda^b \psi \ (P), \ \kappa^b=i\bar\psi  \lambda^b \psi \ (S) 
\label{bilinears}
\end{eqnarray}
with $a=1,2,3$, $b=4,5,6,7$. We denote by $\psi_l$ the light quark $(u,d)$ field doublet in flavour space;  the isospin as well as the pseudoscalar ($P$) or scalar ($S$) character of each operator are indicated, and $\tau^a$ and $\lambda^a$ denote respectively Pauli and Gell-Mann matrices. 

The lowest lying states in the hadron spectrum with the quantum numbers of the above quark operators are the  $f_0(500)$ ($\sigma_l$), mostly dominated by its light quark component,  the light component of the $\eta$ ($\eta_l$), the pion ($\pi^a$), the $a_0(980)$ ($\delta^a$), the kaon ($K^b$) and the $K_0^* (700)$ ($\kappa^b$).

The interest to study the above hadronic states in connection with the QCD transition is that for exact restoration of the chiral symmetry $SU(N_f)_V\times SU (N_f)_A$, with $N_f$ the number of light flavors, the pairs $\pi^a-\sigma_l$, $\delta^a-\eta_l$, $K^b-\kappa^b$ can be connected by the chiral group. In other words, they become chiral partners, so that any observable calculated from their correlators, such as susceptibilities ($p=0$ correlator in Fourier space) or spatial screening masses ($p^0=0$, $\vert\vec{p}\vert\rightarrow 0^+$) would degenerate for two given chiral partners in the regime of exact chiral restoration. That would be the case e.g. for $N_f=2$ massless quarks at the QCD transition temperature $T_c$ \cite{Pisarski:1983ms}. We recall that in the physical case of $N_f=2+1$ massive flavours the transition for $\mu_B=0$ is most likely a crossover at $T_c\simeq $ 155 MeV \cite{Aoki:2009sc,Borsanyi:2010bp,Bazavov:2011nk,Ratti:2018ksb,Bazavov:2019lgz}. In addition, the scalar susceptibility, i.e. the $\sigma_l$ $p=0$ correlator, develops a peak around $T_c$ which is actually one of the more reliable sources for establishing the chiral restoration crossover in the lattice. 

As a bonus, the previous operators provide also information about the possibility of the restoration of the $U(1)_A$ symmetry, which was predicted long time ago as an asymptotic mechanism rather than a sharp transition \cite{Gross:1980br,Shuryak:1993ee,Cohen:1996ng,Lee:1996zy}. The relevant issue here, which is still under study in the community, is whether such $U(1)_A$ restoration takes place close enough to the chiral transition. Should that be the case, it would affect many interesting aspects  of the phase diagram such as the order and universality class of the transition \cite{Pisarski:1983ms,Pelissetto:2013hqa}. There are also phenomenological consequences such as the reduction of the $\eta'$ mass, since its main contribution comes from the $U(1)_A$ axial anomaly \cite{Kapusta:1995ww}. As a matter of fact, Bose-Einstein correlations can be fitted experimentally with  a reduced in-medium $\eta'$ mass \cite{Csorgo:2009pa}. Such decreasing trend is also measured in the lattice \cite{Kotov:2019dby}. 

From the viewpoint of the bilinear operators considered above, now  $\pi^a-\delta^a$, $\sigma-\eta_l$, $K^b-\kappa^b$ can be connected by a $U(1)_A$ rotation and therefore degeneration of observables for those pairs would be the relevant signal to study the strength of $U(1)_A$ breaking at the chiral transition. Note that the $K-\kappa$ degenerate both for chiral and  $U(1)_A$ restoration. 

The results obtained by lattice collaborations regarding degeneration of chiral and $U(1)_A$ partners show some tension among them. On the one hand, $N_f=2$ analyses close to the light chiral limit show compatibility between chiral and $U(1)_A$ restoration and a small gap between them for massive quarks \cite{Cossu:2013uua,Tomiya:2016jwr,Brandt:2016daq}. On the other hand, including strangeness, i.e. for $N_f=2+1$ simulations, and  physical masses,  $U(1)_A$ symmetry is still significatively broken at the chiral transition 
 \cite{Buchoff:2013nra}. As we will see below, our recent analysis based on  Unitarized Effective Theories and Ward Identities  helps to reconcile those results and emphasizes the role of thermal resonances.

 \section{The role of thermal resonances}
 \label{sec:reso}
 
 The spectral properties of resonances develop a dependence with temperature and chemical potentials when those resonances are generated and decay inside the thermal bath or in-medium environment. In some particular cases, such dependence, which can be inferred from hadron effective theories, turns out to be crucial to explain some of the aspects of the QCD phase diagram discussed in the introduction. 
 
 A remarkable example  is the $f_0(500)$ (formerly known as $\sigma$) resonance. The nature and even the existence of that state has been the object of a miriad of studies in the past and is nowadays firmly established as a broad pole in the  second Riemann sheet of the $I=J=0$ $\pi\pi$ scattering amplitude \cite{Berman:1964eq,Garcia-Martin:2011nna,ParticleDataGroup:2020ssz}, which is successfully described as dynamically generated within unitarized effective theories \cite{Pelaez:2015qba,Pelaez:2021dak}. Its thermal dependence has been obtained from a Chiral Perturbation Theory (ChPT) calculation of the unitarized finite-temperature $\pi\pi$ scattering amplitude \cite{GomezNicola:2002tn,Dobado:2002xf} and the connection with chiral symmetry restoration has been established by obtaining the scalar susceptibility through the following saturated approximation \cite{Nicola:2013vma,Ferreres-Sole:2018djq} 
 
 \begin{figure*}
\centering
\includegraphics[width=0.45\textwidth]{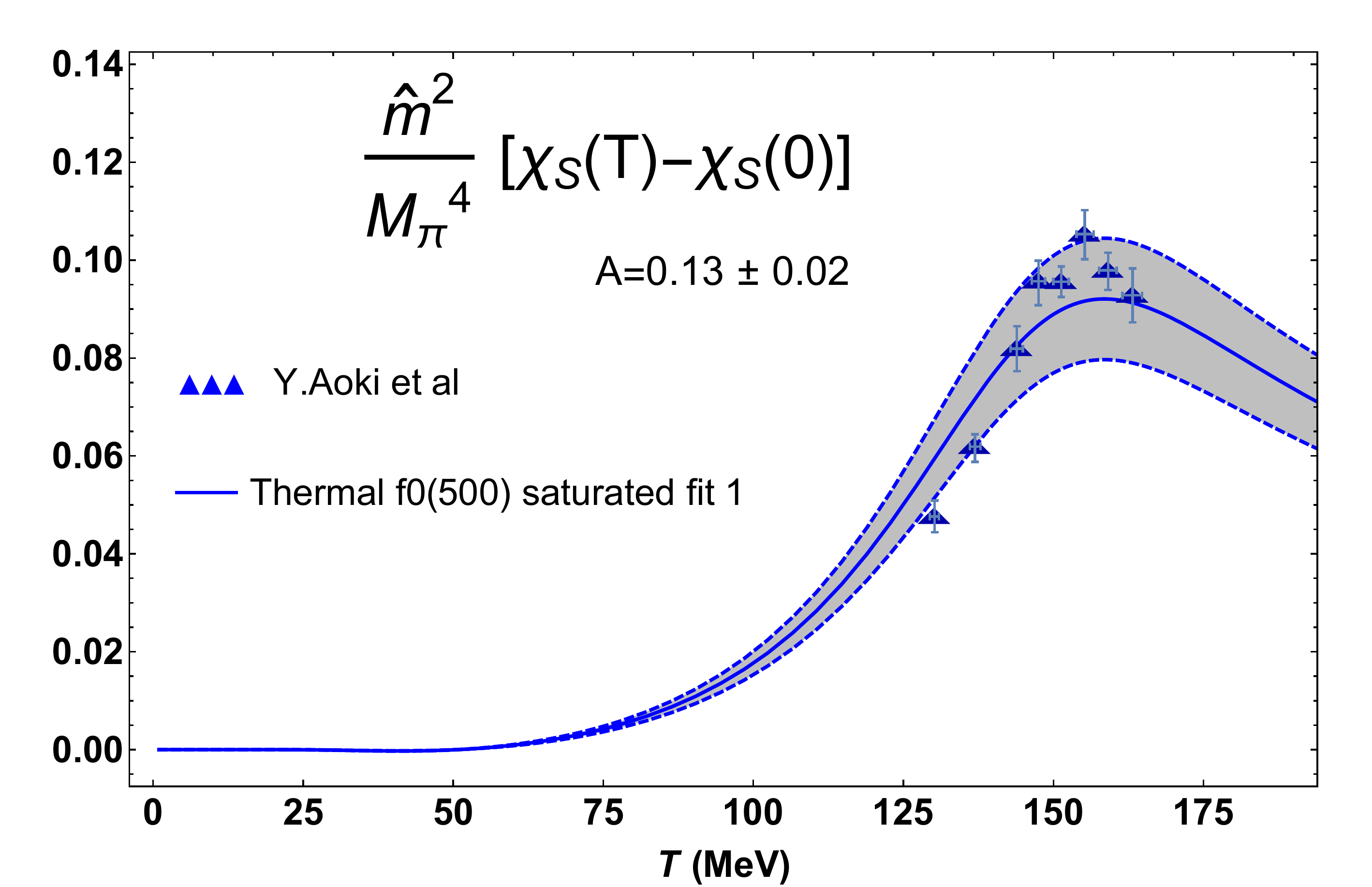}
 \includegraphics[width=0.45\textwidth]{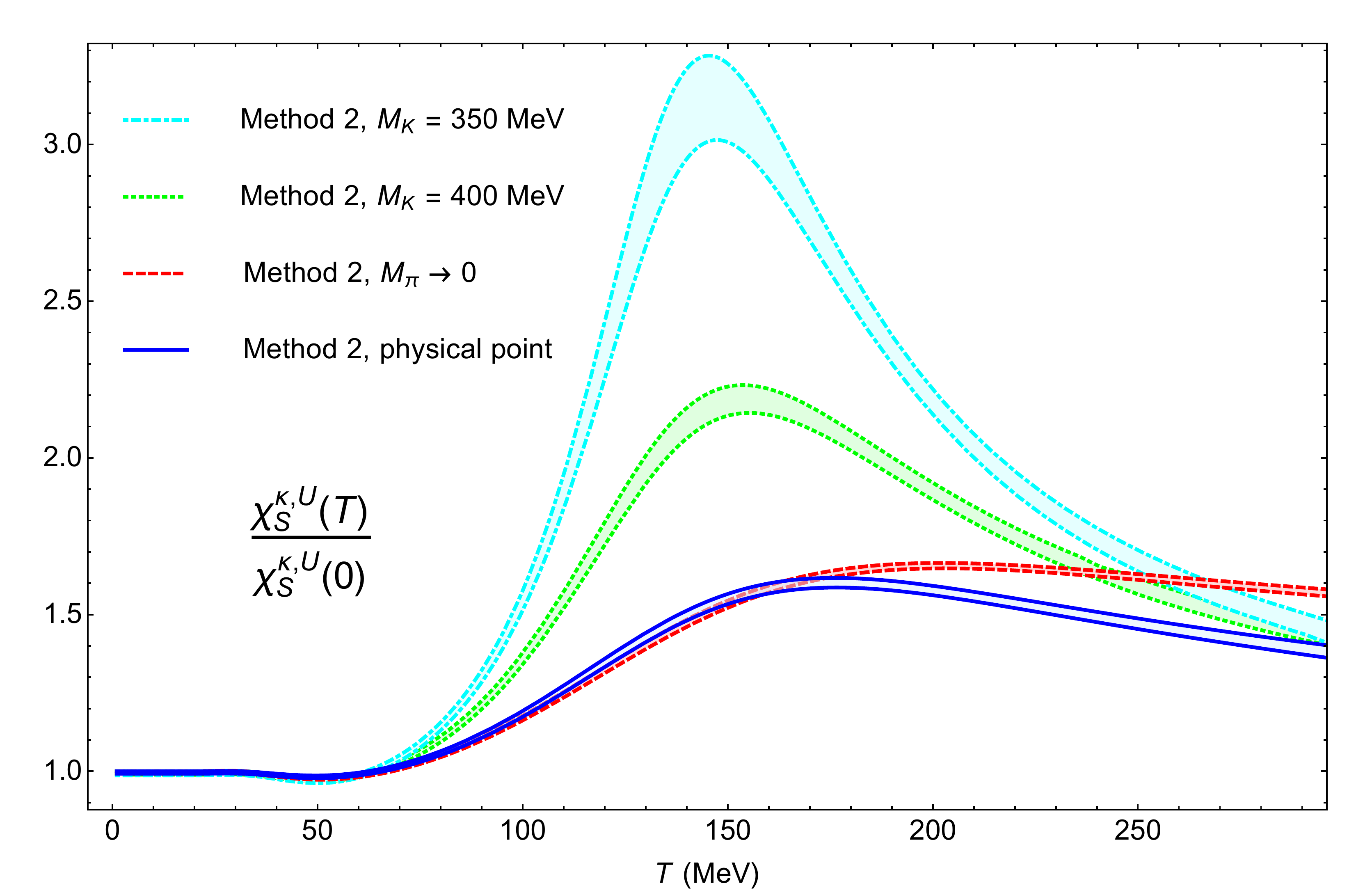}
 \caption{$I=0$ (left) and $I=1/2$ (right) scalar susceptibilities within the saturation approach, from \cite{Ferreres-Sole:2018djq} and \cite{GomezNicola:2020qxo} respectively. In the left panel, the lattice points are from \cite{Aoki:2009sc}, only points up to $T_c\simeq$ 155 MeV are included in the fit and the uncertainty  corresponds to the 95\% confidence level. In the right panel the uncertainty bands correspond to variations within the LEC uncertainty range.} 
\label{fig:resonances}    
\end{figure*}
 
 \begin{equation}
\frac{\chi_S (T)}{\chi_S(0)}\simeq\frac{M_S^2(0)}{M_S^2(T)},
\label{saturatedsus}
\end{equation} 
 where $M_S^2$ is the real part of the resonance pole $s_p=(M-i\Gamma/2)^2$ of the $\pi\pi$ unitarized scattering amplitude, which plays the role of the scalar $f_0(500)$ $p=0$ self-energy \cite{Ferreres-Sole:2018djq}.  In Figure 
 \ref{fig:resonances} (left) we show the result of a fit to lattice data of the saturated susceptibility within the unitarized ChPT approach, as given in \cite{Ferreres-Sole:2018djq}. The fit parameter, denoted $A$ is just a normalization constant, which takes the value $A_{ChPT}=0.15$ when $\chi_S(0)$ is chosen as the scalar susceptibility calculated in ChPT  \cite{GomezNicola:2010tb,GomezNicola:2012uc} with the low-energy constants (LEC) used in \cite{Ferreres-Sole:2018djq}. These results show that the expected peak of the scalar susceptibility around the transition is well reproduced just with the thermal $f_0(500)$ and the fit parameter lies within the ChPT value. Actually,  even without fitting, the LEC uncertainty covers already  the lattice points below $T_c$. In addition, when comparing this unitarized saturated approach with a HRG calculation there is a significant improvement in the former over the latter, which actually fits numerically the lattice points but does not have a peak profile \cite{Ferreres-Sole:2018djq}.

 A similar approach has been followed for the $I=1/2$ scalar susceptibility, i.e., that corresponding to the $\kappa$ channel \cite{GomezNicola:2020qxo}. In that case the relevant process is $\pi K$ scattering at finite temperature and the saturated approach is carried out using \eqref{saturatedsus}, now for the $K_0* (700)$ thermal pole. The result is also given in Figure \ref{fig:resonances} (right) for one of the unitarization methods given in \cite{GomezNicola:2020qxo} (other methods  give similar results).  A very interesting conclusion for this channel is that a susceptibility peak also appears, but now significatively above $T_c$. As we will discuss in section \ref{sec:WI}, the presence of that peak and the behaviour of the susceptibility around is directly linked to the interplay between chiral and $U(1)_A$ restoration, providing direct quantitative information about the role of strangeness and quark masses in that context. Actually, note that as the kaon mass is reduced towards the pion one, the susceptibility peak becomes more pronounced and the peak temperature decreases, approaching $T_c$.  This is nothing but a natural consequence of approaching the regime of degenreated $SU(3)$ flavor symmetry as long as the quark masses are concerned. In such regime, the scalar/pseudoscalar nonet members tend to degenerate. In particular, in that limit the  poles of the scalar $K_0^* (700)$ and the octet component of the $f_0(500)$ become degenerate \cite{Oller:2003vf} and so they do the $\pi$ and $K$ pseudoscalar states \cite{RuizdeElvira:2017aet}. Thus, the susceptibility peak in the $\kappa$ channel inherits naturally the properties of the $\sigma$ channel one as the kaon mass drops. Another interesting limit displayed also in Figure  \ref{fig:resonances} is the light chiral limit, for which, as we will see below, we expect that the $\kappa$-channel peak flattens for temperatures above that peak, due to a more rapid degeneration beween the $\kappa$ and $K$ states. 

 \section{Ward Identities for susceptibilities and condensates}
 \label{sec:WI}
 
 A series of recent works has shown that  Ward Identities (WI) derived formally from the QCD generating functional can be used to relate susceptibilities and quark condensates in different channels, which is indeed quite useful regarding the issues commented above for the  QCD phase diagram  \cite{Nicola:2013vma,Nicola:2016jlj,GomezNicola:2017bhm,Nicola:2018vug,GomezNicola:2020qxo}. In particular, the following identities turn out to be  quite relevant for our present discussion:

\begin{align}
\chi_P^{ls}(T)&=-2\frac{m_l}{m_s} \chi_{5,disc}(T) \label{wils5}\\
=-\frac{2}{m_l m_s}\chi_{top}(T)&=\frac{1}{2m_l m_s} \left[m_l \condl(T)+m_l^2\chi_P^{ll}\right],
\label{witop}\\
\chi_P^K(T)&=-\frac{\condl (T)+2\conds (T)}{m_l + m_s},
  \label{wikaon} \\
  \chi_S^\kappa (T)&=\frac{\condl (T)-2\conds (T)}{m_s-m_l}.
  \label{wikappa} 
\end{align}

Here, $m_l=m_u=m_d$, $\chi_P^{ls}$ is the pseudoscalar crossed $\eta_l\eta_s$ susceptibility with $\eta_s=i\bar s \gamma_5 s$, $\chi_{5,disc}=\frac{1}{4}\left[\chi_P^\pi-\chi_P^{ll}\right]$ with $\chi_P^\pi$, $\chi_P^{ll}$, $\chi_P^K$, $\chi_S^\kappa$ the $\pi\pi$, $\eta_l\eta_l$, $KK$, $\kappa\kappa$ susceptibilities, respectively, and $\chi_{top}=(-1/36)\chi_P^{AA}$ is the   topological susceptibility and $\chi_P^{AA}$ is  the correlator of the anomaly operator $A(x)=\frac{3g^2}{32\pi^2}G_{\mu\nu}^a\tilde G^{\mu\nu}_a$ with $G_{\mu\nu}^a$ the gluon field tensor.

The interest of the  identity \eqref{wils5} lies in the fact that, as discussed above, the $\eta_l$ field can be transformed into  $\delta$ by a chiral $SU(2)_A$  rotation, under which the $\eta_s$ field remains invariant. Therefore, should the chiral symmetry be exact, $\chi_P^{ls}$ would be degenerated with the $p=0$ $\delta\eta$ correlator, but the latter is odd under parity transformations and therefore vanishes from parity conservation. Thus, if we combine such parity argument with the WI given in eq.\eqref{wils5}, one arrives to the conclusion, at least for these channels, that exact chiral symmetry restoration implies exact $U(1)_A$ restoration since $\chi_{5,disc}$ requires {\em both} symmetries to be restored so that $\pi$ and $\eta_l$ degenerate. 

\begin{figure*}
\centering
\includegraphics[width=0.45\textwidth]{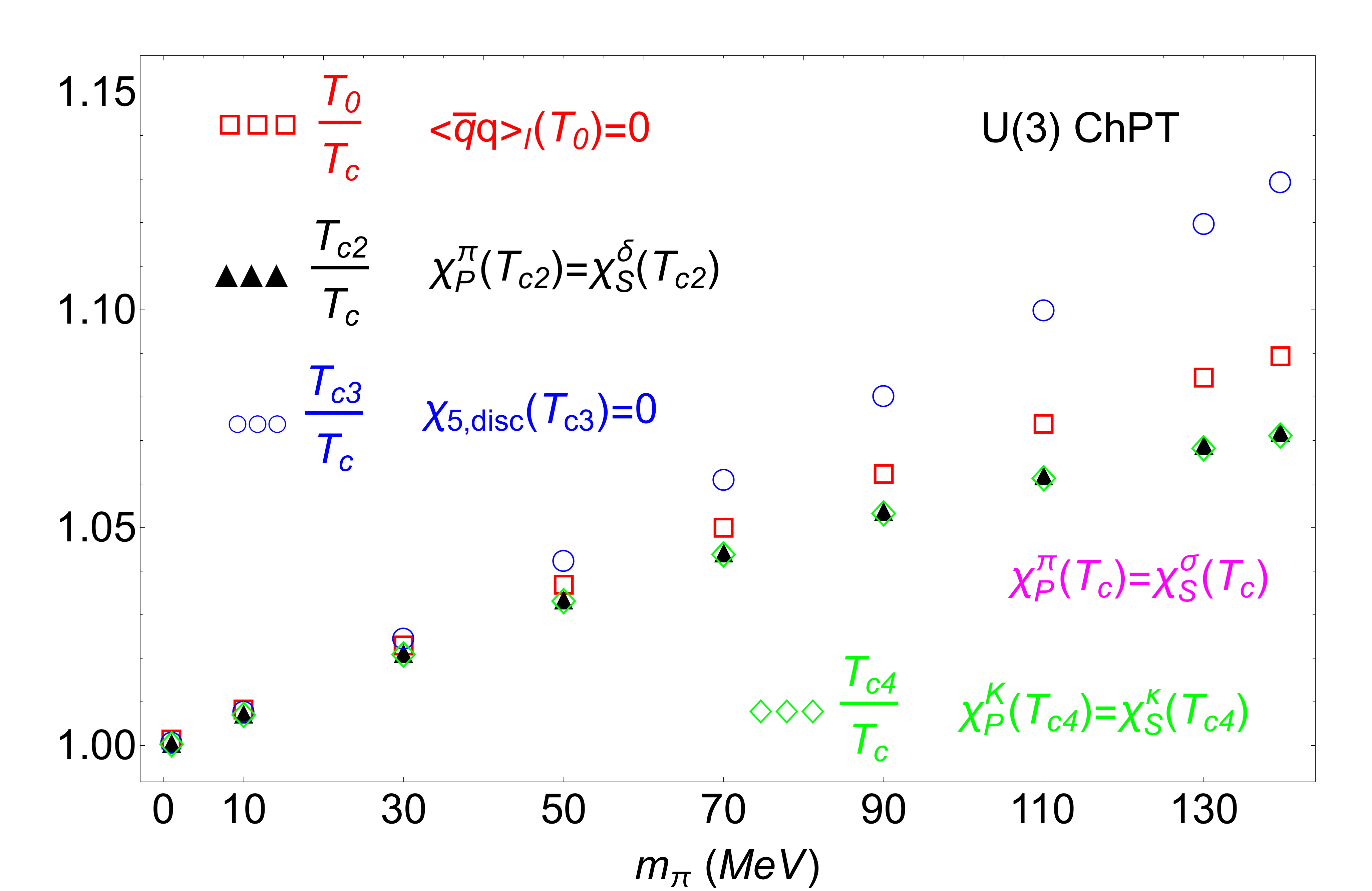}
\includegraphics[width=0.45\textwidth]{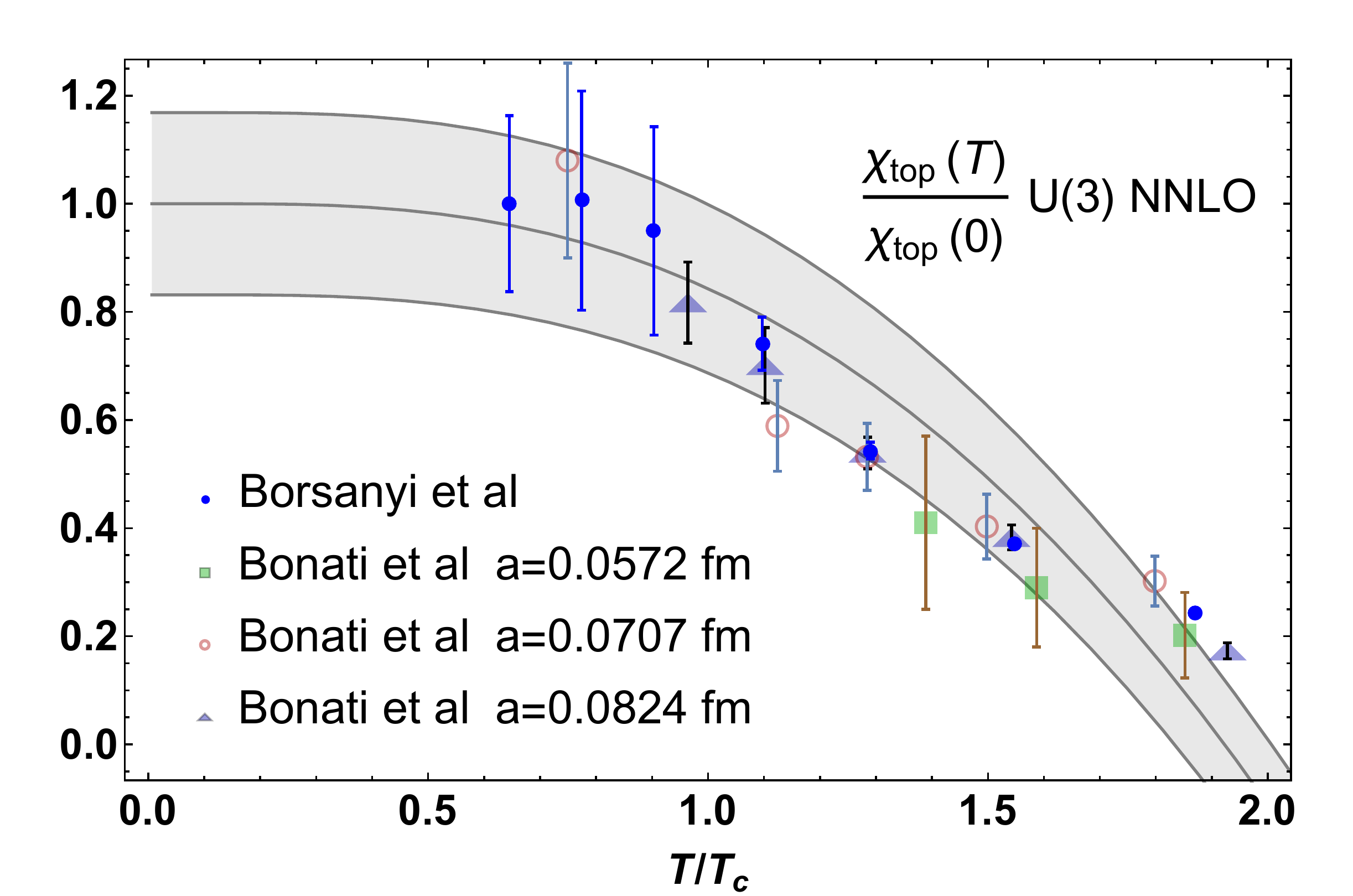}
\includegraphics[width=0.45\textwidth]{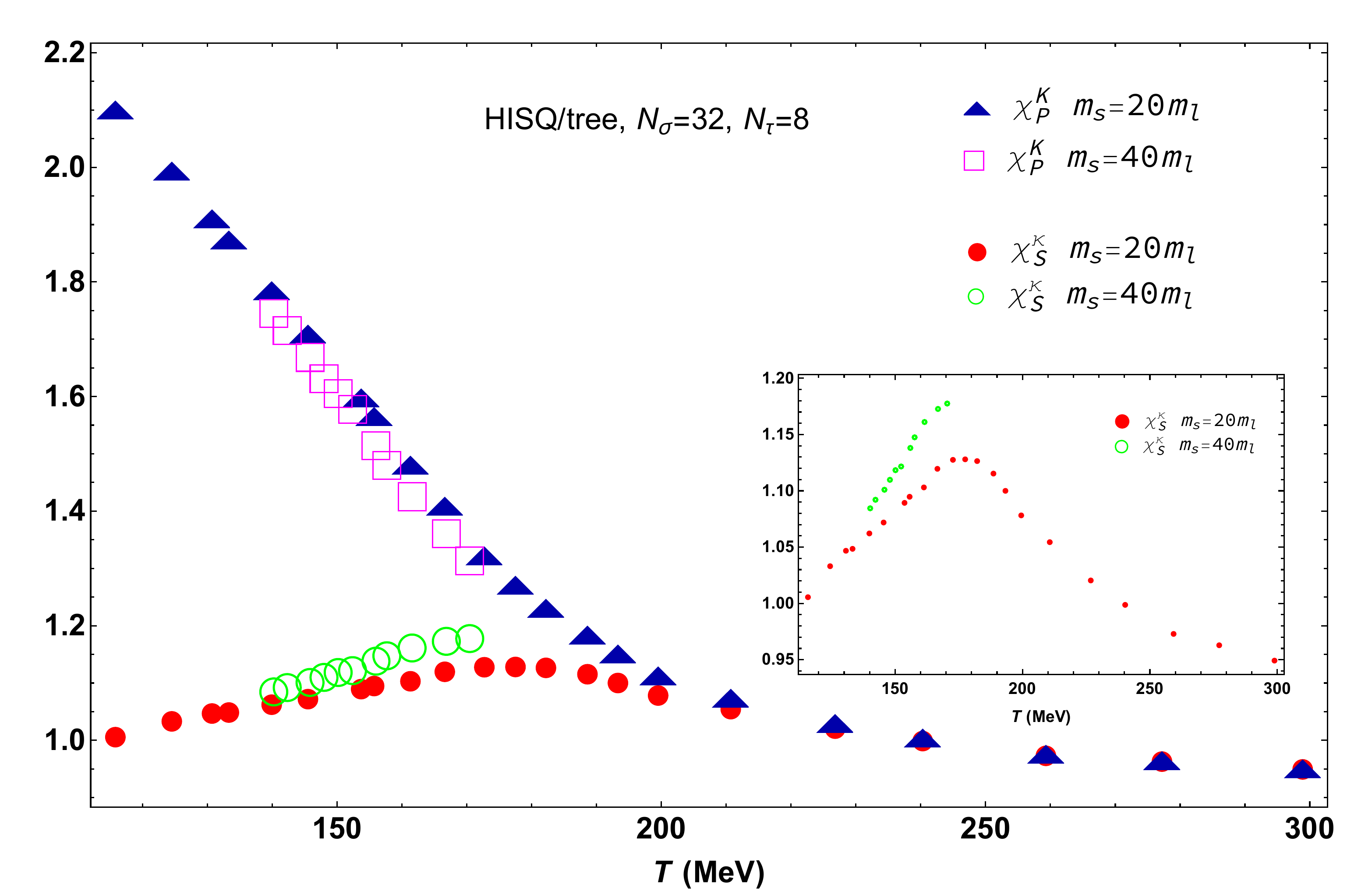}
 \caption{Uppper left panel: Evolution of  different degeneration temperatures with the pion mass, from \cite{Nicola:2018vug}. Upper right panel: Topological susceptibility at finite temperature from  \cite{GomezNicola:2019myi}. The lattice points in the right panel correspond to \cite{Bonati:2015vqz,Borsanyi:2016ksw} while the bands reflect the LEC uncertainty. Lower panel: Reconstructed kaon and kappa susceptibilities, from the analysis in \cite{GomezNicola:2020qxo} with condensate lattice points taken from \cite{Bazavov:2011nk}.} 
\label{fig:chiraltc}    
\end{figure*}

The above  conclusion based on WI is totally consistent with the result obtained in lattice simulations for $N_f=2$, which are compatible with $U(1)_A$ restoration close to $T_c$ as the chiral limit is approached. Confirmation of the above results has also been obtained   within a recent $U(3)$ ChPT calculation \cite{Nicola:2018vug}. In that work, degeneration of $U(1)_A$ partners takes place above the chiral ones for physical meson masses, consistently with lattice analyses. As the light chiral limit is approached by reducing the pion mass, one obtains the result showed in Figure \ref{fig:chiraltc} for the degeneration temperatures of different partners. Namely, we denote by $T_c$ and $T_0$ the chiral restoration temperatures corresponding to $\pi-\sigma$ degeneration and the vanishing of $\condl$, $T_{c2}$ corresponds to the $U(1)_A$ $\pi-\delta$ degeneration,  $T_{c3}$ stands for the chiral and $U(1)_A$ restoration of the $\pi-\eta_l$ partners ($\chi_{5,disc}$ vanishing) while $K-\kappa$ channels degenerate at $T_{c4}$. It becomes pretty clear that all those degeneration temperatures tend to coincide as the light chiral limit is approached, confirming our previous remarks based on WI. On the other hand, a remarkable result, which will help to understand the role of strangeness, is that $K-\kappa$ degeneration almost coincides  with the $\pi-\delta$ one even for the physical pion mass and is therefore linked to $U(1)_A$ degeneration. The latter conclusion has also been confirmed by a PNJL-model analysis of meson screening and pole masses, combined with lattice data \cite{Ishii:2016dln}.

As for the identity \eqref{witop}, it establishes the topological susceptibility as an alternative measure of joint chiral and $U(1)_A$ restoration, providing a neat separation into one contribution proportional to the light quark condensate  plus another one where the $\eta_l$ susceptibility enters. The first contribution is responsible for chiral restoration and dominates for low temperatures and up to the chiral transition, while the second one accounts mostly for the residual $U(1)_A$ breaking above the chiral transition. That behaviour has indeed been reproduced also within $U(3)$ ChPT in 
 \cite{GomezNicola:2019myi}. The advantage of that formalism is that it incorporates naturally the effect of $\eta'$ loops, which turn out to be of the same order as those of $\pi,K,\eta$ ones. Actually, the dominant contribution to the topological susceptibility, both at zero and finite temperature, comes from the lightest states, i.e., the pions, as it can be seen already from  the leading order ChPT contribution (temperature independent)  
 
 \begin{equation}
\chi_{top}^{U(3),LO}= -\frac{1}{2}\condl^0 \frac{M_0^2 {\bar m}}{M_0^2+6B_0{\bar m}} 
\label{chitoploib}
\end{equation}
where  $\condl^0$ is the quark condensate in the chiral limit, $M_0$  stands for the anomalous part of the $\eta'$ mass and $\displaystyle\bar m^{-1}=\sum_{i=u,d,s} m_i^{-1}$ which shows clearly the dominance of light states.  Also for that reason the finite-temperature analysis in this formalism captures reasonably well results from the lattice, even for temperatures well above the ChPT applicability range, as it is shown in Figure \ref{fig:chiraltc}.

Finally, identities \eqref{wikaon}-\eqref{wikappa} open up the possibility of considering the $K-\kappa$ sector for the study of the strangeness effect in the interplay between chiral and $U(1)_A$ restoration and explain also the behaviour found for the susceptibility in the kappa channel from the thermal resonance method discussed in section \ref{sec:reso}. Thus, those identities predict a constantly decreasing $\chi_P^K$ since both light and strange condensates decrease, but the minus relative sign between those two condensates in \eqref{wikappa} implies that the initially  increasing behaviour of $\chi_S^\kappa$ with temperature dominated by the decreasing of (negative) $\condl$is taken over by the strange condensate at some point above the chiral transition, from which $\chi_S^\kappa$ would start decreasing  tending to degenerate with $\chi_P^K$. Therefore, this argument explains the peak obtained for $\chi_S^\kappa$ in section \ref{sec:reso} and, together with our previous WI and effective theory arguments, suggests that such peak and its behaviour above it is an indication of how $U(1)_A$ is restored above the chiral transition, quantifying  the role of strangeness through $\conds$. The previous argument is further supported by the lattice, as Figure \ref{fig:chiraltc} shows, where the kaon and kappa susceptibilities are reconstructed using identities \eqref{wikaon}-\eqref{wikappa} and the quark condensates obtained in the lattice collaboration \cite{Bazavov:2011nk}. In particular, from our previous arguments, as the light chiral limit is approached we  expect 
a more rapid growth just below the peak, from chiral restoration, and a flattening just above it from the  faster $U(1)_A$ $K-\kappa$ degeneration.

\section{Conclusions}

Light thermal resonances $f_0(500)$ and $K_0^* (700)$ are key to understand chiral and $U(1)_A$ restoration. Combining  effective theory approaches with Ward Identities  support $U(1)_A$ restoration for fully restored chiral symmetry and allows one to understand the role of quark masses and strangeness in the physical regime of massive quarks and $N_f=2+1$ flavors. 

\section*{Acknowledgments}
 Work partially supported by research contract  PID2019-106080GB-C21 ( ``Ministerio de Ciencia e Innovaci\'on"),  the European Union Horizon 2020 research and innovation program under grant agreement No 824093 and the Swiss National Science Foundation, project No.\ PZ00P2\_174228. A. V-R acknowledges support from a fellowship of the UCM predoctoral program.

%\end{multicols}
%\medline
%\begin{multicols}{2}

%\end{multicols}
\end{document}